# The Study of a Temperature Dependence of the Dark Rate of Single Electrons Emitted from a Cathode of a Multi-Cathode Counter as a Method to Search for Hidden Photons of CDM.

A.V.Kopylov, I.V.Orekhov and V.V.Petukhov

Institute of Nuclear Research of Russian Academy of Sciences, 117312 Prospect of 60[th] Anniversary of October Revolution 7a, Moscow, Russian Federation

The dark rate of single electrons emitted from a metal cathode of the counter depends upon a temperature. At cryogenic temperatures the dark rate measured with bi-alkaline cathodes of PMT has been found to increase by lowering the temperature of the detector. One of the possible interpretations of this effect suggests the energy trapping mechanism from the conversion of Hidden Photons of CDM at the surface of a photocathode. The measurements are planned by using a Multi-Cathode Counter (MCC) with a copper and aluminum cathodes of 0.2 m$^2$ at temperatures from 300 K down to 220 K to spot the increase of cryogenic dark rate from Hidden Photons of CDM. The physical motivation and the description of the experiment are presented.

Hidden photons (HPs) were proposed by L.B.Okun in 1982 [1]. They can be observed through a kinetic mixing term $(\chi/2)F_{\mu\nu}X^{\mu\nu}$ with the ordinary photons [2], where $\chi$ is a dimensionless parameter quantifying the kinetic mixing. Here $F_{\mu\nu}$ is the field stress of the ordinary electromagnetic field $A^\mu$ and $X^{\mu\nu}$ is the field stress of the HP field $X^\mu$. Electromagnetic field **A** can be considered as accompanying the field of hidden photons **X** so that a dark matter solution for HP with a mass $m_{\gamma'}$ can be described as [3]

$$\begin{pmatrix} \mathbf{A} \\ \mathbf{X} \end{pmatrix}\bigg|_{DM} = \mathbf{X}_{DM} \begin{pmatrix} -\chi \\ 1 \end{pmatrix} \exp(-i\omega t) \qquad (1)$$

i.e. has a spatially constant mode k = 0, oscillating with frequency $\omega = m_{\gamma'}$. A new method with a spherical mirror has been proposed in [3]. In this method ordinary photons which accompany hidden photons with frequency $\omega = m_{\gamma'}$ are emitted in the direction perpendicular to the surface of the mirror. These photons can be detected by PMT placed in the center of the mirror sphere. This has been performed in [4] where an upper limit 6x10$^{-12}$ has been obtained for a mixing parameter $\chi$ for the hidden photon mass $m_{\gamma'}$ = 3.1 ± 1.2 eV. The method has principal limitation: by higher frequencies the mirror losses the reflectivity with the resulting drop in sensitivity. It is worth to note that in this method the target for hidden photons is a surface of the mirror and the effect is proportional to the surface of the mirror. The power collected by the mirror in this case is taken to be [3]

$$P = 2\alpha^2\chi^2\rho_{CDM}A_{dish} \qquad (2)$$

Where: $\alpha^2 = \cos^2\theta$, $\theta$ is the angle between the HP field, when the direction of the field is the same everywhere, and the plane of a mirror, and $\alpha^2 = 2/3$ if HPs have random orientation; $\rho_{CDM} \approx 0.3$ GeV/cm$^3$ – is the energy density of CDM which is taken here to be equal to the energy density of HPs and $A_{dish}$ – the mirror's surface. It means that here the part $2\chi^2$ of the energy of the field of HPs gets converted at the surface of the mirror into the light. The experiment tests exactly this process whatever be the mechanism of the conversion. The natural question arises: what would be for higher frequencies? It is known that when a metal sheet gets irradiated by UV photons and the energy of UV photons is greater than a work function of the metal, the single electrons are emitted from the surface of the metal. Utilizing this effect a new device has been developed – a Multi-Cathode Counter (MCC) for the detection of single electrons emitted from metal cathode as a result of the conversion of HPs into ordinary photons. The device was constructed, tested and used for the search of HPs [5,6]. The dark rate $R_{MCC}$ of single electrons emitted from the metal cathode has been measured. The power collected by a cathode of a counter can be expressed through $R_{MCC}$:

$$P = \frac{R_{MCC} \cdot m_{\gamma'}}{\eta} \qquad (3)$$

Here $\eta$ – quantum efficiency which was taken here equal to the one for ordinary photon of the frequency $\omega = m_{\gamma'}$. Combining the expressions (2) and (3) one can get the sensitivity of this method

$$\chi_{sens} = 2.9 \times 10^{-12} \left(\frac{R_{MCC}}{\eta\ 1\ \text{Hz}}\right)^{½} \left(\frac{m_{\gamma'}}{1\ \text{eV}}\right)^{½} \left(\frac{0.3\ GeV/cm^3}{\rho_{CDM}}\right)^{½} \left(\frac{1\ m^2}{A_{MCC}}\right)^{½} \left(\frac{\sqrt{2/3}}{\alpha}\right) \qquad (4)$$

From the measured value of $R_{MCC}$ the upper limit for a mixing parameter $\chi < 10^{-9}$ has been set [6] for the masses of HPs from about 5 eV till 1 keV with a minimal value of about $7 \cdot 10^{-10}$ at $m_{\gamma'} = 15$ eV.

The measurements of dark rates at various temperatures were conducted in experiments [7,8] with photomultipliers. In the work of H.O.Meyer [8] the measurements were performed till the lowest temperature 4 K and it was shown that when the temperature is lowered the dark rate decreases according to Richardson's law but below 220 K the dark rate starts increasing again. This "cryogenic" dark rate is well described by the expression

$$I = G\ A\ \exp(-T/T_c) \qquad (5)$$

Where $G \approx 5$ cm$^{-2}$ s$^{-1}$; A – the surface of a cathode of PMT; $T_c \approx 100$ K. The remarkable feature of this cryogenic dark rate is that electrons are emitted in bursts which are distributed randomly in time but the events within a burst are correlated. As the temperature decreases, the rate of bursts as well as the number of events per burst, increase. The observed time distributions are indicative of a trap mechanism. This feature 'somewhat burstlike and non statistical in character' of non thermal dark current has been observed also in an earlier work by J.P.Rodman and H.J.Smith [7]. The observed phenomenon had no physics explanation. The data in [8] have been obtained with photomultipliers with bi-alkali cathodes (K-Cs-Sb) on a thin platinum backing. But cryogenic emission has been observed also with multi-alkali (Sb-Na-K-Cs) cathodes, and without a metal backing [7]. Bi-alkali and especially multi-alkali cathodes have a low work function. Approximation by Richardson's law of the curve of the variation of a dark rate as a function of reciprocal temperature, see Fig.16 in [9] for bi-alkali cathode, shows that $\varphi_W \approx 0.4$ eV. This value is in agreement with the temperature dependence of thermionic dark rate measured in [8]. It would be very interesting to get the data for metal cathodes with higher work function, for example, made of platinum ($\varphi_W \approx 5.4$ eV) or copper ($\varphi_W \approx 4.4$ eV). What would be the parameters G and $T_c$ in the expression (5) in this case? To describe this burst-like behavior at cryogenic temperatures one can suggest a following scheme. The electrons in the burst are emitted from some "hot" spots which act like traps for the energy absorbed by a cathode. It was shown in [7] that detailed scan of photocathodes reveals fine structure in quantum efficiency of different areas of photocathodes which could explain the origin of these "hot" spots. When the energy accumulated in these traps overcomes a certain threshold level the energy gets released from the trap by a burst of electrons. The electrons are emitted in a string of events and the interval between the single electron emissions gets increased upon decreasing of the energy accumulated in a trap. When the energy accumulated by trap gets smaller than a work function the emission of electrons terminates. The competing process for the accumulation of energy in the trap is a thermal radiation. The lower is a temperature, the weaker is a thermal radiation and the higher is the probability that energy accumulated in a trap overcomes a certain threshold level. For a metal with a high work function the probability to overcome the threshold is much less than for metal with a low work function because of the competing process of thermal radiation. One can anticipate that the parameters G and $T_c$ in expression (5) for copper both are smaller than for bi-alkali cathode. If the source of energy is the photons from the conversion HP-photon absorbed by a cathode of PMT then quantum efficiency is also temperature dependent value because only a portion $\varepsilon(T) = \exp(-T/T_c)$ of energy absorbed by a cathode gets converted into electrons emitted. So the quantum efficiency $\eta(T) = \varepsilon(T) \cdot m_{\gamma'}/\varphi_W$. At zero K all energy absorbed by a cathode is converted into the electrons emitted and the quantum efficiency is maximal $\eta(0\ K) = m_{\gamma'}/\varphi_W$. Then it is easy to accommodate the expression (4) to this case of cryogenic emission:

$$\chi = 2.9 \times 10^{-10} \left( G \cdot \varphi_W \right)^{1/2} \left( \frac{0.3\ GeV/cm^3}{\rho_{CDM}} \right)^{1/2} \left( \frac{\sqrt{2/3}}{\alpha} \right) \qquad (6)$$

The remarkable feature of this expression is that it does not contain a mass of hidden photon. If we put in this expression G = 5.0 cm$^{-2}$s$^{-1}$ from [8] and take $\varphi_W \approx 0.4$ eV we obtain $\chi \approx 4.1 \cdot 10^{-10}$ for $\rho_{CDM}$ = 0.3 GeV/cm$^3$ and $\alpha = \sqrt{2/3}$. Thus from the measurements of [8] at a temperature 4 K we have found a mixing parameter and it turned out to be a huge one. In fact, this is only a lower estimate because of the uncertainty on the reflectance of the cathode for a photon of frequency $\omega = m_{\gamma'}$ which is unknown. Certainly, this result obtained from the measurements of [8] can be considered at most only as a hint for possible effect from hidden photons. It definitely needs a further study. The measurements with platinum or copper or any other metal with relatively high work function would be helpful to clear this issue. This is principally important to make measurements with different metals because the very trapping mechanism belongs to a solid state physics and is very dependent upon the technology of the fabrication of a cathode. So what is observed with one cathode can not be observed with another one. To make sure that one does not omit the effect the measurements should be done with different cathodes, only the data collected with different metal cathodes can provide complete information to understand the real mechanism. The crucial point is what would be observed at 4K, where the temperature dependent term $\varepsilon(4\ K) \approx 1$. The gaseous detectors do not work at this temperature. The lowest temperature which can be reached with the argon-methane mixture is of about 220 K. So it's impossible to measure dark rate at 4K by a multi cathode counter. But still the measurements at 220 K may be very illuminating. If we assume that parameter $G^{(4\ K)}$ at 4 K is inversely proportional to a work function, then comparison of dark rates measured for copper cathode at a temperature T directly yields $T_c$ for a copper cathode provided the expression (5) is valid for any metal the cathode is made of.

$$T_c^{(Cu)} = (T - 4\ K) / \ln(G^{(4\ K)}_{bi-alk}\varphi_{W1} / R^{(T)}_{MCC}\varphi_{W2}) \qquad (7)$$

Here $\varphi_{W1}$ and $\varphi_{W2}$ are the work functions of bi-alkaline and copper cathodes, $G^{(4\ K)}_{bi-alk} = G = 5.0$ cm$^{-2}$s$^{-1}$ and $R^{(T)}_{MCC}$ is a dark rate measured by a multi cathode counter with a copper cathode at a temperature T. Figure 1 shows what can be obtained in these measurements with a multi-cathode counter in the temperature range from 300 K down to 220 K. Solid lines (approximation) and crosses (experimental points) are taken from Fig.1 of [8]. The dashed line at Fig.1 shows the rates corresponding to expression (5) which can be observed by our detector if to take $T_c \approx 30$ K to match with the upper limits of the rate obtained as a very preliminary result (magenta points) in our current measurements at 300 and 307 K. At present the measurements are continued to increase the accuracy. Large surface of a cathode of MCC and the possibility to measure the rates in two different configurations of MCC with similar geometries enable to make the measurements five orders of magnitude below the thermionic rate of bi-alkaline cathode. The point at 4 K for MCC was taken to be $G^{(4\ K)}_{bi-alk} \varphi_{W1} / \varphi_{W2}$. The quantum efficiencies were taken here: $\varphi_{W1} = 0.4$ eV and $\varphi_{W2} = 4.4$ eV. The latter value corresponds to atomically clean copper. In our case a routinely cleaned copper is used for a cathode, so the real value of it can be somewhat lower what means that the point at 4 K for MCC can be somewhat higher than is depicted at Fig. 1. The magenta line shows the rates expected in the region from 300 K down to 220 K where the measurements are planned with a multi-cathode counter. One can see that they can be very helpful to spot the increase of cryogenic dark rate if it really takes place. We hope that future development of the MCC technique will enable to increase the accuracy of measurements till the level adequate to trace the effect along magenta line. The yellow points show the rates expected at a temperature below 220 K what is beyond the access for MCC.

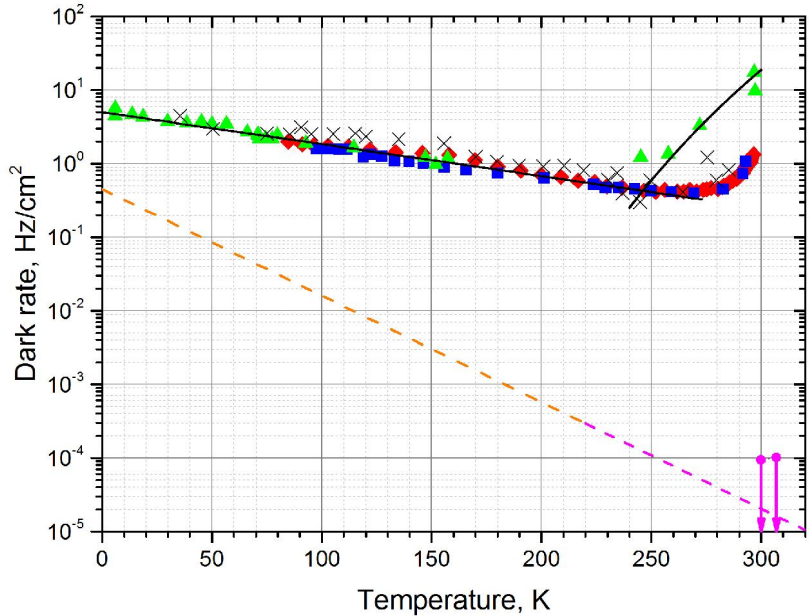

Figure 1. Dark rates for bi-alkaline (upper curve) and copper (lower curve) cathodes at different temperatures.

**Conclusions**

The increase of the cryogenic dark rated observed in several experiments [7,8] can be interpreted as a manifestation of hidden photons of CDM. To test this hypothesis is necessary to make the measurements with a metal cathode with higher work function. The study of a temperature dependence of the dark rate of single electrons emitted from a metal cathode can be very helpful for the search of hidden photons of CDM. Presently developed Multi-Cathode Counter (MCC) [5,6] with a copper cathode and a new counter with an aluminum cathode will be used to make the measurements from 300 K down to 220 K. The aim is to spot the possible increase of cryogenic dark rate as a manifestation of the effect from hidden photons of CDM and to reevaluate the mixing parameter $\chi$ which is found here to be $\chi \approx 4.1 \cdot 10^{-10}$ from the measured cryogenic rate at 4 K in [8].

**Acknowledgements**